\documentclass[preprint,cite]{epl}
\usepackage{amsmath,amssymb}
\title{Equilibrium Onions?}
\author{L. Ramos\inst{1} \and D. Roux\inst{2} \and
  P.~D.~Olmsted\inst{3} \and M.~E.~Cates \inst{4} }
\shortauthor{Ramos \etal}
\institute{
 \inst{1} Groupe de Dynamique des Phases Condens\'{e}es
(UMR CNRS-UM2 n$^o$5581), CC26, Universit\'{e} Montpellier 2,
34095 Montpellier Cedex 5, France\\
  \inst{2} Centre de Recherche Paul Pascal, avenue Schweitzer, 33600 Pessac,
  France\\
  \inst{3}  Department of Physics and Astronomy \& Polymer IRC, University
  of Leeds, Leeds LS2 9JT, UK \\
  \inst{4}  School of Physics,
University of Edinburgh,
JCMB Kings Buildings,
Mayfield Road, Edinburgh EH9 3JZ
Scotland}
 {Surfaces and interfaces: phase transitions}
 {Lyotropic phases}
\begin{document}
\maketitle
\begin{abstract}
  We demonstrate the possibility of a stable equilibrium
  multi-lamellar (``onion'') phase in pure lamellar systems (no excess
  solvent) due to a sufficiently negative Gaussian curvature modulus.
  The onion phase is stabilized by non-linear elastic moduli coupled
  to a polydisperse size distribution (Apollonian packing) to allow
  space-filling without appreciable elastic distortion. This model is
  compared to experiments on copolymer-decorated lamellar surfactant
  systems, with reasonable qualitative agreement.
\end{abstract}
\section{Introduction}
\label{sec:introduction}

The existence of vesicles at thermal equilibrium is a
long-standing, and still controversial, problem \cite{laughlin97}.
Unilamellar and multi-lamellar vesicles (MLVs) were first induced
in lyotropic lamellar phases with excess water by adding energy
(\textit{e.g.} shear flow, ultrasound, electric field)
\cite{Papahadjopoulos}.  In a very elegant work, using membranes
generated by a chemical reaction, Hoffman and co-workers
demonstrated that any of lamellae, unilamellar vesicles, or MLVs
can be prepared in the same system, depending on the mechanical
path chosen \cite{Hoffmann}. However in some special cases
\cite{kaler89,herve93} \textit{equilibrium} unilamellar vesicles
have been demonstrated. These systems are all in the dilute regime
(large excess of water).  Some experiments have suggested that
MLVs (or onions) can be stabilized in the semi-dilute regime
(excess water) \cite{herve93}. Theories to explain the stability
of dilute unilamellar vesicles either describe a competition
between the entropy of mixing and the curvature energy of the
vesicles \cite{israelachvili,herve93,BKL94} or a symmetry-breaking
instability leading to a spontaneous curvature \cite{Safr+91b}.
Onions are also predicted to be stabilized in the dilute and
semi-dilute regimes due to an unstable curvature energy; in these
cases, a transition towards unilamellar vesicles is avoided by
imposing either a core energy \cite{Fournier.Durand94} or a cutoff
in the entropy of the Helfrich interactions \cite{simons92}.
Indeed, Simons and Cates \cite{simons92} have discussed the
stability of unilamellar vesicles and the transition to onions as
the concentration increases even when the curvature energy is
unfavorable, due to entropic reasons.

In the concentrated regime (homogeneous lamellar phase with no
excess water), very monodisperse onions can be prepared under
shear \cite{diat93a,diat93,SierRoux97}. The vesicles fill space
and remain in the one phase region without expelling solvent, by
distorting into space-filling polyhedra \cite{Guli+96}.  The
resulting texture is a lattice of disclinations which can be
either disordered \cite{diat93a} or ordered \cite{SierRoux97}.
Moreover many experimental systems ``spontaneously'' exhibit
onions in regions of the phase diagrams that seem to be
continuously linked to the lamellar phase \cite{Augu+97}. This
evolution from a texture of polydisperse vesicles to a texture of
focal conics upon tuning a parameter, such as a co-surfactant
concentration or the volume fraction of solvent, seems to be
generic. A striking signature of this evolution is the change in
the rheology, from a viscous gel in the ``onion'' phase to a more
fluid phase in the ``lamellar'' focal conic phase. Thus, whether
an onion texture is at equilibrium or not seems yet unclear.  In
this work we propose a model based on a non-quadratic elastic
energy, which shows that a pure lamellar phase, or one of
space-filling deformed monodisperse onions, is unstable with
respect to a polydisperse space-filling packing (we study an
Apollonian packing \cite{bork94}).  We compare these predictions
to experiments on copolymer-doped lamellar phases, originally
introduced by Ligoure \textit{et al.}  \cite{Ligoure99}.
\section{Model}
\label{sec:model}
The bending free energy per area $F/A$ for membranes is usually taken,
following Helfrich, as the simplest quadratic function of the mean and
Gaussian curvatures, respectively $2H=r_1^{-1}+r_2^{-1}$ and
$G=1/(r_1r_2)$, where $r_1$ and $r_2$ are the two principal radii of
curvature \cite{helfrich73}. This suffices for large curvature radii
$r\gg\delta$, where $\delta$ is the layer thickness. However, if
onions exist, because of the high concentration of membrane, high
curvatures (small $r$) are expected near the core of the onions. We
will thus use the free energy, relative to a flat state, of a
symmetric bilayer expanded to quartic order in the curvature radii
\cite{fournier97a}:
\begin{align}
\frac{F}{A} & = 2\kappa\, H^2 +\bar{\kappa}\, G + \tfrac14{c_1}H^4
+ \tfrac14{c_2}G^2 + 2{c_3}G H^2,
\end{align}
where $\kappa$ and $\bar{\kappa}$ are the conventional bending and
Gaussian curvature moduli, and the non-linear moduli $c_i$ can be
expected to scale as $k_BT \delta^2$.

Stability of a flat membrane requires $2\kappa>-\bar{\kappa}>0$. For
$\bar{\kappa}>0$ an instability towards a phase with $G<0$
(\textit{e.g.} a bicontinuous phase) is possible.  However, for
$\tilde{\kappa}\equiv2\kappa+\bar{\kappa}<0$ spherical shells are
stabilized, with free energy (for radius $r$)
$F_{\textrm{shell}}(r)<0$ given by
\begin{equation}
  \label{eq:2}
  F_{\textrm{shell}}(r) = 4\pi\tilde{\kappa} +  \frac{\pi\tilde{c}}{r^2},
\end{equation}
where $\tilde{c} = c_1+4\,c_2+ c_3$.  Positive $\tilde{c}$ stabilizes
spherical shells (vesicles) of finite radius. Unilamellar vesicles
cannot accommodate volume fractions close to space filling, so we
consider the free energy of an onion of radius $R$, constructed from
$k$ shells of discrete radii $r_j=jd,\, (j=1,2,3,\ldots,k)$ set by the
layer spacing $d$, which we assume remains fixed:
\begin{align}
  F_{\mathrm{onion}}(R=kd) &= \sum_{j=1}^{k} F_{\rm shell}(r_j=jd) =
    \frac{4\pi\tilde{c}}{d^2}\left[\lambda k +
      S(-2,k)\right]
    \label{eq:4}
\end{align}
where $\lambda\equiv\tilde{\kappa}d^2/\tilde{c}$ is the balance
between the Helfrich and non-linear elastic coefficients and
$S(l,m)\equiv \sum_{i=1}^m i^l$\footnote{There is generally, a surface
  energy due to, \textit{e.g.}, van der Waals attraction, but if we
  assume an inter-onion spacing identical to the equivalent lamellar
  separation we need not consider such a term.}.
In the limit $R\gg d$ the second term above, due to the
fourth order coefficients, is equivalent to the core energy introduced
by Fournier and Durand \cite{Fournier.Durand94}.

\begin{figure}
  \begin{center}
    {\includegraphics[scale=0.8]{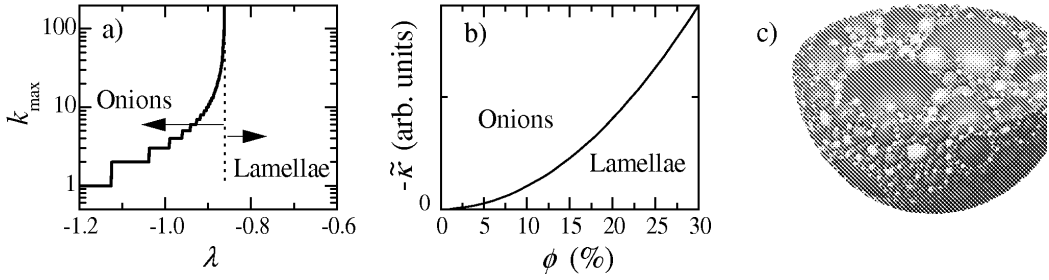}}
    \caption{(a) Maximum number of layers per onion, $k_{\mathrm{max}}$,
      as a function of $\lambda\equiv\tilde{\kappa}d^2/\tilde{c}$.
      Apollonian onions are stable for $-1.125\lesssim\lambda\lesssim
      -0.86$. (b) Schematic phase diagram in the
      $(\tilde{\kappa},\phi)$ plane,
      $\tilde{\kappa}^{\ast}\sim\phi^2$. (c) Schematic packing of the
      Apollonian distribution (reprinted with permission from
      \cite{bork94}).}
    \label{fig:app1}
  \end{center}
\end{figure}

For large dilute monodisperse onions the free energy per unit volume
scales as $F/V\sim F_{\rm onion}(R)\sim\tilde{c}(\lambda +
d/R)/(d^3R^2)$, leading for $\lambda<0$ to stable onions with size
$R\sim d/|\lambda|$.  However, in the concentrated regime monodisperse
onions cannot fill space without non-trivial deformations
\cite{Guli+96}. Typically, onions formed under shear pack as
polyhedra, with the deformation of layer $j$ concentrated along the
edges of length $r_j$ with curvatures of order $1/\xi$, where the
smectic bending penetration length $\xi=\sqrt{K/\bar{B}}$ depends on
the bending $K=\kappa/d$ and compression $\bar{B}$ moduli of the
lamellar phase \cite{Pani+96}.  Summing over all shells in a deformed
polyhedron leads to an energy per unit volume that scales as
$F_{el}/V\sim \tilde{c}[\lambda + d^2/\xi^2]/(R \xi d^3)$. For
$\lambda+d^2/\xi^2<0$ this leads to an unphysical unilamellar vesicle
phase with $R=d$ (the lower cutoff).  However, a polydisperse phase
can relax the deformation, leading, for a broad enough distribution to
a space-filling distribution of spheres. A candidate space-filling
distribution is the Apollonian packing, for which the number
distribution of spheres $n(R)\sim (R/d)^{-D_A}$, with a fractal
dimension estimated from simulations to be $D_A\simeq 3.45$
\cite{anishchik95,Rouault99,bork94} (see Fig.~\ref{fig:app1}c).  Such
a packing has been previously suggested to describe focal conic
defects in smectic droplets \cite{bidaux73}. In principal we should
calculate this distribution from a free energy that incorporates
entropy.  Although small on the typical scale of the elastic terms,
this entropy should widen the parameter range in which onions are
stable \cite{simons92}.  Existing equations of state for,
\textit{e.g.} polydisperse hard spheres, fail for widely disperse
distributions near space filling \cite{Blaak.Cuesta01}, so we have
been unable to estimate this contribution to any useful accuracy.
Entropic effects will most likely influence the upper and lower
cutoffs of the distribution, and would eventually lead (for highly
fluctuating systems) to an exponential, rather than power law, size
distribution away from space filling \cite{BKL94,Fournier.Durand94}.

The normalized number distribution of onion sizes $R_k$ obeying
$\tfrac{4\pi}{3}\sum_{k=1}^{k_{max}}n(R_k)  R_k^3=V_m$, where
 $V_m$ is the volume of lamellar material, is
\begin{equation}
  \label{eq:6}
  \tilde{n}(k=R_k/d) = \frac{3}{4\pi  d^3}\frac{V_m}{S(3-D_A,k_{\mathrm{max}})}
\left(\frac{R_k}{d}\right)^{-D_A}.
\end{equation}
The total energy of the distribution, $F_{tot} =
\sum_{k=1}^{k_{\mit max}} \tilde{n}(k) F_{\rm onion}(k)$, is
\begin{align}
  \frac{4\pi d^3F_{tot}}{3 V_m} &=\frac{
  \lambda\,S(1-D_A,k_{\mathrm{max}}) +
  \sum_{j=1}^{k_{\mathrm{max}}} S(-2,j)j^{-D_A}}
  {S(3-D_A,k_{\mathrm{max}})}.
\end{align}
This energy can then be minimized over the maximum onion size
$k_{\mathrm{max}}$ as a function of the single parameter $\lambda$ to
find the lowest free energy distribution (Fig.~\ref{fig:app1}a).
Although spherical shells are stable for $\lambda<0$, the stabilizing
elastic constants $c_i$ assure that a finite and negative
$\lambda\leq\lambda^{\ast}\simeq-0.86$ is required for onion
formation. Onions are first stable with `infinite radius', implying a
smooth transition from lamellae to onions. However, the scaling
$nR^3\sim R^{3-D_A}\simeq R^{-0.45}$ implies an unbounded total
volume. Hence, the size distribution must have an upper cutoff, which
in practice should be the smallest dimension of the sample container.
(In fact, the nature of the size distribution for Apollonian packings
of spheres is still under debate \cite{anishchik95,Rouault99}, and
appears to depend crucially on the manner in which the packing is
constructed.)  The maximum onion size then decreases, and for
$\lambda\lesssim-1.125$ the stable phase is of unilamellar vesicles.
At this point the calculation has presumably broken down, and either
another phase such as micelles intervenes\footnote{A cubic phase of
  unilamellar vesicles has been experimentally observed in the
  concentrated regime \cite{Gradzielski99}. }, or even higher order
curvature terms become important. Because $\lambda=\tilde{\kappa}
d^2/\tilde{c}$, and the volume fraction $\phi$ is of order $\delta/d$,
the condition $\lambda < \lambda^{\ast}$ implies a phase boundary
$\tilde{\kappa}^{\ast}\sim\phi^2$, as shown in Fig.~\ref{fig:app1}b,
in the $\tilde{\kappa}\!-\!\phi$ plane.

\begin{figure}
  \begin{center}
     \includegraphics[scale=0.6]{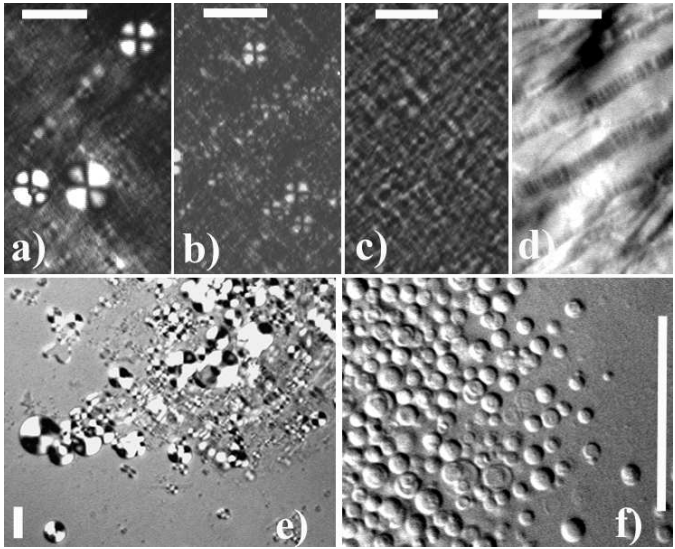}
    \caption{Top:  Gel phase observed between crossed
      polarizers, for membrane fractions $\phi$ and
      copolymer/surfactant weight ratios $\alpha$ of (a) $\phi=16\%$,
      $\alpha=0.2$; (b) $\phi=16\%$, $\alpha=0.4$; (c) $\phi=16\%$,
      $\alpha=0.8$; (d) $\phi=22\%$, $\alpha=0.8$.  Bottom:
      Gel phase in contact with water; (e, crossed polarizers)
      $\phi=16\%$, $\alpha=0.15$; (f, Differential Interference
      Constrast microscopy) $\phi=16\%$, $\alpha=1.6$.  Scale bar
      $50\,\mu\mathrm{m}$.}
    \label{fig:pictures}
  \end{center}
\end{figure}

\section{Experiments}
\label{sec:exp}
As shown theoretically in \cite{PortLigo95}, adsorption of amphiphilic
copolymers to surfactant bilayers is expected to decrease
$\tilde{\kappa}$, which eventually becomes negative. We therefore
prepare lamellar phases decorated with copolymers \cite{Ligoure99}.
Bilayers of thickness $\delta_{0}=2.8 \, \rm{nm}$ comprising
cetylpyridinium chloride (CpCl) and octanol (Oct)
($\rm{CpCl/Oct}=0.95$ w/w) are diluted in brine ($\rm{[NaCl]}= 0.2
\rm{M}$) and decorated with Symperonics F68 (Serva)
[$\rm{(EO)_{76}}-\rm{(PO)_{29}}-\rm{(EO)_{76}}$, where EO is ethylene
oxide and PO is propylene oxide]) as amphiphilic copolymer. The
bilayer volume fraction $\phi$ and the copolymer/surfactant weight
ratio $\alpha$ range from $9$ to $23\%$ and $0$ to $1.6$,
respectively. The bilayer volume fraction $\phi$ and the
copolymer/surfactant weight ratio $\alpha$ range from $9$ to $23\%$
and $0$ to $1.6$, respectively. The lamellar structure is preserved
upon adding copolymer, with a continuous hardening into a ``lamellar
gel''. The location of this gel in the ($\alpha$,$\phi$) phase diagram
was studied by Ligoure \textit{et al.} \cite{Ligoure99}. They observed
gel-like behavior over a range of critical copolymer/surfactant weight
ratios that depends on the membrane fraction.

\begin{figure}
  \begin{center}
        \includegraphics[scale=0.8]{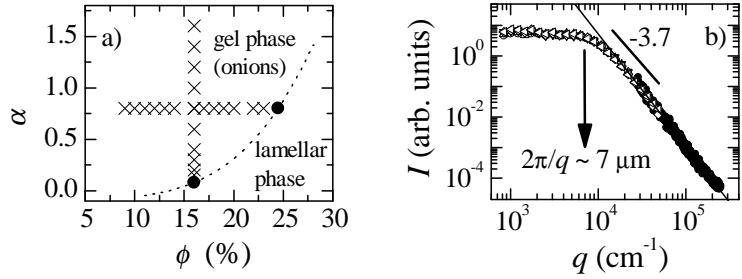}
  \end{center}
    \caption{(a) Schematic phase diagram, adapted from
      \cite{Ligoure99}, overlaid with the experimental points
      ($\times$), where $\alpha$ and $\phi$ are respectively the
      copolymer/surfactant weight ratio and bilayer volume
      fraction.  Experimental phase boundaries are inferred at
      ($\bullet$). The dotted line is a guide for the eye. (b) Small-
      and large-angle static light scattering intensity for
      $\phi=16\%$ and $\alpha=0.8$. A crossover from power law to
      constant intensity is noted at a length scale of order
      $7\,\mu\textrm{m}$.  }
    \label{fig:scatter}
\end{figure}

The marked variation of the mechanical properties of the bulk lamellar
phase is accompanied by modifications of the optical textures. The
pictures shown in Fig.~\ref{fig:pictures} demonstrate that the
lamellar gels are onion phases. Onions are clearly obtained for both
neat lamellar phases and samples put in contact with solvent. In the
latter case individual onions detach from the solvent/gel interface
(Fig.~\ref{fig:pictures}e,f), while in the former case defect textures
characteristic of the onion phase are observed between crossed
polarizers (Fig.~\ref{fig:pictures}a-c). The onions are highly
polydisperse, with an apparent maximum size which seems to decrease
with increasing $\alpha$.  Different textures and the presence of oily
streaks are discerned at higher $\phi$ (Fig.~\ref{fig:pictures}d).

\begin{figure}
  \begin{center}
    {\includegraphics[scale=0.8]{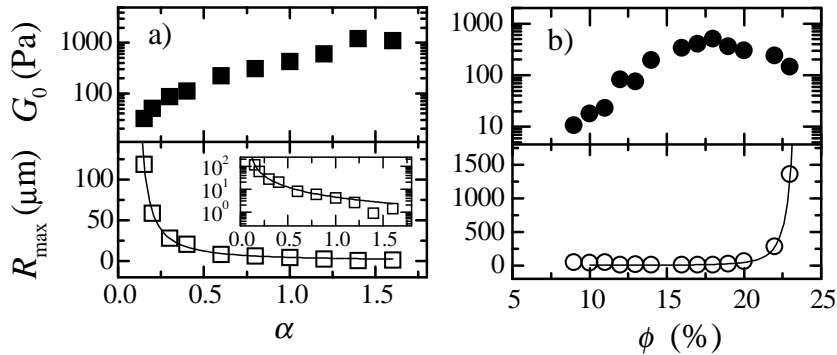}}
    \caption{Variation of the elastic plateau modulus $G_0$ (top) and
      maximum radius $R_{max}$ (bottom) determined from $G_0$, for samples with
      (a) $\phi=16\%$ and various $\alpha$; (b) $\alpha=0.8$ and
      various $\phi$. Solid lines are fits to $R\sim
      (\alpha-\alpha_c)^{-p}$ and $R\sim (\phi_c-\phi)^{-m}$, with
      $\alpha_c=0.08, \phi_c=24.5\%, p=1.3, m=3.0$. (Inset, a) same data as in the main
      figure in a semi-logarithmic plot.}
    \label{fig:moduli}
  \end{center}
\end{figure}

While microscopy indicates that the size distribution is very broad
and varies with $\alpha$ and $\phi$, a quantitative determination of
the size distribution using this technique is delicate. However,
assuming an Apollonian distribution we can, for example, estimate the
macroscopic elastic modulus as a mass (or volume) average of the
elastic moduli of onions of different sizes $r_i$.  The elastic energy
stored in a linear deformation of strain $\mu$ of an onion of radius
$R$ is $4\pi\gamma R^2\mu^2$, where the effective surface tension
$\gamma=\tfrac12\sqrt{K\bar{B}}$ \cite{Linden93}, leading to a modulus
$G_R=3\pi\gamma/R$.  Thus, an estimate of the volume-averaged elastic
modulus is $G_0\simeq \frac1V\sum_i 3\gamma n_i r_i^{2} $
\cite{Pani+96}.  Averaging for $R_{max}\gg d=\delta_{0}/\phi$, we find
$G_0\simeq\tfrac{3\gamma}{d}
\tfrac{4-D_A}{D_A-3}\left(\frac{d}{R_{max}} \right)^{4-D_A} \simeq\
3.67 \, \gamma \, d^{-0.45} \, R_{max}^{-0.55}$ for $D_A=3.45$. The
linear elastic plateau is measured in a Couette rheometer along two
lines in the phase diagram at constant $\alpha$ and $\phi$
(Fig.~\ref{fig:scatter}a). For fixed $\phi$, the modulus $G_0$
increases monotonically with increasing $\alpha$, while for fixed
$\alpha$, the modulus varies non-monotonically with $\phi$
(Fig.~\ref{fig:moduli}). To extract the maximum onion radius $R_{max}$
from $G_0$ we must estimate the surface tension $\gamma$. For
interlamellar forces dominated by Helfrich entropic undulations,
$\sqrt{K\bar{B}}=\frac{3\pi}{8}\frac{k_{B}T}{(d-\delta)^2}$, with
$\delta$ the bilayer thickness \cite{Ligoure99,RouxSafinya88}.
Interestingly, it has been shown \cite{Ligoure99} that for a decorated
lamellar phase the inter-bilayer interactions are still of an
effective Helfrich form, with an effective bilayer thickness
$\delta_{\it eff}=\delta_{0}+2 \times h_{pol}$ where $h_{pol}$ is the
apparent thickness of the polymer layer \footnote{Since the lamellar
  phase is stabilized by Helfrich interactions, one should in
  principle incorporate layer undulations into an onion entropy
  \cite{simons92}.  However, we believe that these effects, important
  for small onions in dilute systems, would not significantly change
  the phase diagram at the higher concentrations studied here.}. In
the experimental range of $\alpha$ we take $h_{pol}=A \alpha ^{1/3}$
with $A=3.5 \, \rm{nm}$, although this expression is strictly valid
only in the brush regime ($\alpha \geq 0.5$). One can thus calculate
the surface tension by taking $\delta=\delta_{\it eff}$ and hence
extract, from the experimental values of $G_{0}$, the variations of
$R_{max}$ with the two experimental parameters $\phi$ and $\alpha$.
As can be seen in Fig.~\ref{fig:moduli}, for $\phi=16\%$, $R_{max}$
varies between $1\,\mu\textrm{m}$ and $120\,\mu\textrm{m}$, with an
apparent divergence near $\alpha_c\simeq0.08$; while for $\alpha=0.8$,
$R_{max}$ varies between $4\,\mu\textrm{m}$ and $1400\,\mu\textrm{m}$,
with an apparent divergence near $\phi_c\simeq24.5\%$ \footnote{We
  note that, at small $\phi$, $R_{max}$ is a decreasing function of
  $\phi$, which may be be due to the nearby lamellar-to-vesicle phase
  boundary \cite{Ligoure99}.}.

The critical-like behavior of the onion maximum size inferred
experimentally is remarkably similar, qualitatively, to the prediction
of Fig.~\ref{fig:app1}. From the observed critical-like variations in
$R_{max}$ we have identified two points on the boundary between
lamellar and onion phases, in the $(\alpha,\phi)$ plane
(Fig.~\ref{fig:scatter}a).  Using the relation between
$\tilde{\kappa}$ and the parameter $\lambda$ we can estimate
experimental values for $\tilde{\kappa}$. The onion/lamellae phase
boundary is given by
$\lambda^{\ast}=\frac{\tilde{\kappa}^{\ast}d^{2}}{\tilde{c}}=-0.86$
with $\tilde{c}=k_{B}T\delta_{\it eff}^{2}$. For the two
copolymer/surfactant weight ratios $\alpha$ for which the boundary has
been determined (Fig.~\ref{fig:app1}c), we obtain $\tilde{\kappa}=-0.1
k_{B}T$ for $\alpha=0.08$ and $\tilde{\kappa}=-0.6 k_{B}T$ for
$\alpha=0.8$. In agreement with theoretical expectations
\cite{PortLigo95}, $\tilde{\kappa}$ is found to be negative, of order
$1\, k_{B}T$, and to decrease with increasing weight ratio. Consistent
with the theoretical phase diagram (Fig.~\ref{fig:app1}b) and with
previous experimental observations \cite{Ligoure99}, we also find that
the critical value of $\tilde{\kappa}$, below which the onion phase is
stable, decreases when $\phi$ increases.

The radii $R_{max}$ estimated from the elastic moduli are similar to
those observed by  microscopy (Fig.~\ref{fig:pictures}e,f).
For $\alpha=0.15$ the largest observable onion is
$R\simeq35\,\mu\textrm{m}$ and the calculated
$R_{max}(G_0)=120\,\mu\textrm{m}$ (Fig.~\ref{fig:pictures}e), and for
$\alpha=1.6$, the largest observable onion is
$R\simeq5\,\mu\textrm{m}$ (in Fig.~\ref{fig:pictures}f) and the
calculated $R_{max}(G_0)=1.4\,\mu\textrm{m}$. We can also compare
$R_{max}$ to static light scattering experiments
(Fig.~\ref{fig:scatter}). The scattering intensity $I(q)$ scales as a
power law  for large $q$, $I(q) \simeq q^{-3.7}$, and has a plateau for
smaller $q$, indicating that there is no structure at lengths larger
than the cross-over length $2\pi/q\simeq 7\,\mu\textrm{m}$.  This
provides an estimate for the maximum onion size $R_{max}$ that is
very close to the value obtained from $G_{0}$ ($R_{max}=6
\,\mu\textrm{m}$).  While similar power law behavior for high $q$ has
previously been observed for monodisperse onions obtained by shear,
the behavior at large $q$ \cite{Leng} was very different, with a peak
characteristic of the onions size observed in that case. The flat
intensity obtained here clearly indicates the intrinsic polydispersity
of the onions.  Figures~\ref{fig:pictures}e,f also show wide size
distributions.

In summary, we have presented experimental evidence for an equilibrium
phase of polydisperse multi-lamellar vesicles (onions), and
rationalized this by combining a non-linear Helfrich elastic theory
with an assumed polydisperse distribution of onions. Future work
should incorporate the entropy and calculate the nature of the
space-filling distribution.
\acknowledgments We thank the Kavli Institute for Theoretical
Physics at the University of Santa Barbara, where part of this
work was initiated. LR thanks C. Ligoure for fruitful discussions
and L. Cipelletti for help in the light scattering experiments,
and PDO thanks CRPP for hospitality.


\end{document}